\documentclass{ws-procs975x65}
\usepackage{dsfont}

\newcommand{\dd}{\mathrm{d}}
\newcommand{\ee}{\mathrm{e}}
\newcommand{\ii}{\mathrm{i}}
\newcommand{\E}{\mathcal E}
\newcommand{\R}{\mathds R}
\newcommand{\C}{\mathds C}

\def\Xint#1{\mathchoice
   {\XXint\displaystyle\textstyle{#1}}%
   {\XXint\textstyle\scriptstyle{#1}}%
   {\XXint\scriptstyle\scriptscriptstyle{#1}}%
   {\XXint\scriptscriptstyle\scriptscriptstyle{#1}}%
   \!\int}
\def\XXint#1#2#3{{\setbox0=\hbox{$#1{#2#3}{\int}$}
     \vcenter{\hbox{$#2#3$}}\kern-.5\wd0}}

\def\dashint{\Xint-}

\begin{document}

\title{Gowdy-Symmetric Vacuum and Electrovacuum Solutions}

\author{J\"org Hennig} 

\address{Department of Mathematics and Statistics,
           University of Otago,\\
           P.O.~Box 56, Dunedin 9054, New Zealand\\
         E-mail: jhennig@maths.otago.ac.nz}

\begin{abstract}
``Smooth Gowdy-symmetric generalized Taub-NUT solutions'' are a class of inhomogeneous cosmological vacuum models with a past and a future Cauchy horizon. In this proceedings contribution, we present families of exact solutions within that class, which contain the Taub solution as a special case, and discuss their properties. In particular, we show that, for a special choice of the parameters, the solutions have a curvature singularity with directional behaviour. For other parameter choices, the maximal globally hyperbolic region is singularity-free. We also construct extensions through the Cauchy horizons and analyse the causal structure of the solutions. Finally, we discuss the generalization from vacuum to electrovacuum and present an exact family of solutions for that case.
\end{abstract}


\bodymatter


\section{Introduction \label{sec:intro}}

Gowdy-symmetric solutions (i.e.\ solutions with two spacelike Killing vectors) to Einstein's field equations are often studied as useful test cases for investigations of key issues in cosmology, like strong cosmic censorship, spikes, or the BKL conjecture.

A particular class of Gowdy-symmetric vacuum models with spatial three-sphere topology has been introduced in Ref.~\refcite{BeyerHennig2012}: the \emph{smooth Gowdy-symmetric generalized Taub-NUT (SGGTN) solutions}. 
Exact solution are not known for the general case, but we will discuss particular families of exact solutions below. In the general case, existence has been shown with abstract methods. 
It turns out that one can choose two free functions,
which then specify a unique SGGTN solution. 
Each SGGTN spacetime has a past Cauchy horizon, and the two free functions describe the behaviour of the spacetime geometry as this horizon is approached. 
Moreover, with the exception of particular ``singular cases'', these models are generally expected to develop a regular second, future Cauchy horizon.

In terms of a time coordinate $t$ and spatial coordinates $\theta$, $\rho_1$,  $\rho_2$, the metric can be written as
\begin{equation}\label{eq:metric}
  \dd s^2=\ee^M(-\dd t^2+\dd\theta^2)+R_0\left[\sin^2\!t\,\ee^u (\dd\rho_1+Q \dd\rho_2)^2+\sin^2\!\theta\,\ee^{-u} \dd\rho_2^2\right],
\end{equation}
where $R_0>0$ is a constant and $u$, $Q$ and $M$ are functions of $t$ and $\theta$ alone.

Existence of these solutions with a Cauchy horizon at $t=0$ is by no means guaranteed a priori, but can be established as follows. First one can prove \emph{local} existence near $t=0$ with the Fuchsian methods developed in Refs.~\refcite{Ames2013a,Ames2013b}. In a next step, \emph{global} existence in the time interval $0<t<\pi$ follows from a result due to Chru\'sciel\cite{Chrusciel1990}.
Note that the line element \eqref{eq:metric} degenerates at $t=0$ and at $t=\pi$. The degeneracy at $t=0$ can be removed with a coordinate change and comes from a coordinate singularity at the Cauchy horizon. However, since the global existence result only applies for $t<\pi$, the question remains of what happens at $t=\pi$?
This can be answered by studying the behaviour of the metric at $t=\pi$  with methods from soliton theory. For that purpose, note that the essential part of the Gowdy-symmetric Einstein vacuum equations is equivalent to the  Ernst equation
\begin{equation}\label{eq:EE}
 \Re(\E)\cdot \left(-\E_{,tt}-\cot t\,\E_{,t}
        +\E_{,\theta\theta}+\cot\theta\,\E_{,\theta}\right)
 =-\E_{,t}^{\ 2}+\E_{,\theta}^{\ 2}
\end{equation} 
for the complex Ernst potential $\E=f+\ii b$.
Interestingly, this equation belongs to the remarkable class of integrable equations, i.e.\ there is an associated linear matrix problem which is equivalent to the original nonlinear equation via its integrability condition. It is possible to integrate this linear problem along the boundaries $t=0,\pi$ and $\theta=0,\pi$, which allows one to find an explicit formula for the Ernst potential 
at $t=\pi$ in terms of the data at $t=0$, so that the metric at $t=\pi$ can be studied. 
These investigations indicate that 
SGGTN solutions develop a second, future Cauchy at $t=\pi$, except in special cases in which curvature singularities form. These ``singular cases'' occur when 
the imaginary part $b$ of the initial Ernst potential satisfies
 \begin{equation}\label{eq:singular}
  b(t=0,\theta=\pi)-b(t=0,\theta=0)= \pm4.
 \end{equation}
Then the solutions have a curvature singularity at $t=\pi$, $\theta=0$  or at $t=\pi$, $\theta=\pi$.

This is as far as we can get in the general case with abstract methods. 
In order to study further properties in more detail, it is useful to construct and investigate exact solutions.

\section{Three Families of Vacuum Solutions}

We intend to solve initial value problems for the Ernst equation, where we prescribe the Ernst potential at $t=0$. This can be done with a method due to Sibgatullin\cite{Sibgatullin}, which, however, was originally developed to construct \emph{axisymetric and stationary} spacetimes (with a spacelike and a \emph{timelike} Killing vector) and not Gowdy-symmetric solutions. But we can use the formal coordinate transformation 
$\rho  = \ii \sin t\sin\theta$, $\zeta = \cos t\cos\theta$
to coordinates ($\rho,\zeta,\rho_1,\rho_2$) in which the metric \eqref{eq:metric} takes the Weyl-Lewis-Papapetrou form for axisymmetric and stationary spacetimes. Consequently, we can apply Sibgatullin's method to our problem.

For a given initial Ernst potential at $t=0$ (corresponding to $\rho=0$), we consider the analytic continuations
$e(\xi):=\E(\rho=0,\zeta=\xi)$, $\tilde e(\xi):=\overline{e(\bar\xi)}$ for $\xi\in\C$.
Then we solve the integral equation
\begin{equation}\label{eq:inteq}
 \dashint_{-1}^1\frac{\mu(\xi;\rho,\zeta)[e(\xi)+\tilde e(\eta)]\,\dd\sigma}{(\sigma-\tau)\sqrt{1-\sigma^2}}=0
\end{equation}
for $\mu(\xi;\rho,\zeta)$ subject to the constraint
\begin{equation}\label{eq:constraint}
 \int_{-1}^1\frac{\mu(\xi;\rho,\zeta)\,\dd\sigma}{\sqrt{1-\sigma^2}}=\pi.
\end{equation}
Here, $\dashint$ denotes the principal value integral and $\xi:=\zeta+\ii\rho\sigma$, $\eta:=\zeta+\ii\rho\tau$ with $\sigma,\tau\in[-1,1]$. The corresponding Ernst potential can then be obtained from
\begin{equation}\label{eq:EP}
 \E(\rho,\zeta)=\frac{1}{\pi}\int_{-1}^1\frac{e(\xi)\mu(\xi)\dd\sigma}{\sqrt{1-\sigma^2}}.
\end{equation}
In the next subsections we present three families of solutions that have been obtained with this method. ``Solution 1'' and ``solution 2'' below are two different 3-parameter families of solutions. ``Solution 3'' is a more general 4-parameter family that contains the first two families as special cases.

\subsection{Solution 1}
We consider the  initial Ernst potential\footnote{The initial Ernst potential must satisfy some conditions at the axes $\theta=0$ and $\theta=\pi$, which explains the special form of the cubic function here.}
\begin{equation}\label{eq:indat1}
 t=0:\quad \E=c_1(1-x^2)
      +\ii x\left[c_3(x^2-3)-x\right],\quad x:=\cos\theta,
 \quad
 c_1>0,\quad c_3\in\R.
\end{equation}
The corresponding initial value problem for the Ernst equation was solved with Sibgatullin's method in Ref.~\refcite{BeyerHennig2014}. The resulting solution depends on the three parameters $c_1$, $c_3$ and $R_0$. We do not give the explicit solution here, since it can be obtained from the metric for ``solution 3'' below in the limit $d\to\infty$. 

First we note that the Taub solution\cite{Taub51} is contained as the special case $c_3=0$. 
In that particular case, the solution is spatially homogeneous, but otherwise we have inhomogeneous cosmological models.

Generally, the solution is regular in the entire ``Gowdy square'' $(t,\theta)\in[0,\pi]\times[0,\pi]$. Only in the singular cases \eqref{eq:singular}, which correspond to $c_3=\pm1$, the Kretschmann scalar $K$ blows up at $t=\pi$, $\theta=0$ or $t=\pi$, $\theta=\pi$. Interestingly, we observe a directional behaviour of this divergence: Depending on the curve along which the singular point is approached, $K$ either tends to $+\infty$, to $-\infty$, or even to any finite value.
This is similar to the behaviour of the Curzon solution\cite{Curzon1925}, where it turned out that it is actually possible to extend the solution beyond the singularity, see Refs.~\refcite{ScottSzekeres1986a,ScottSzekeres1986b}. The Curzon singularity was later classified as a \emph{directional singularity}. In our case, however, despite the directional behaviour of $K$, there is no directional singularity in the strict sense. Instead, along each curve towards the singularity where $K$ remains finite, other curvature invariants diverge.

We can construct several extensions of the solutions through the Cauchy horizons. In terms of $x=\cos\theta$ and $y=\cos t$, the solution is initially defined for $(x,y)\in[-1,1]\times[-1,1]$. However, after a suitable transformation of the coordinates $\rho_1$, $\rho_2$, it is possible to extend the solution to all $y\in\R$, $x\in[-1,1]$.

Finally, one can show that the past extensions always contain a curvature singularity, and the future extensions contain either one or two curvature singularities, depending on the parameter values.

\subsection{Solution 2}
Another solution with quite different properties can be obtained from the initial Ernst potential
\begin{equation}\label{eq:indat2}
t=0:\quad \E =c_1(1-x^2)\left(1-\frac{x}{d}\right)-\ii x^2,\quad
c_1>0,\quad d>1.
\end{equation}
Together with $R_0$ we again obtain a 3-parameter family of solutions. The application of Sibgatullin's method to this problem is described in Ref.~\refcite{Hennig2014}. For the metric we can once more refer to the more general ``solution 3'' below, which reduces to this solution in the limit $c_3\to0$. Now we discuss some properties of these spacetimes.

The solutions are always regular in the Gowdy square, since
there is no parameter choice for which the singularity condition \eqref{eq:singular} is satisfied.

Whereas ``solution 1'' has a past Cauchy horizon with \emph{closed} null-generators, the past Cauchy horizon of this solution is generally ruled by \emph{non-closed} null generators.

The causal structure of this solution turns out to be more complex than the one of ``solution 1''. Again we can construct (several) extensions through the past and future Cauchy horizons into the regions with $y\in\R$, $x\in[-1,1]$. These extensions contain further Cauchy horizons through which the solution can be extended 
(again there are several extensions) 
into coordinate domains with $|x|>1$. The metric in these regions is singular along a certain hyperbola. But this turns out to be a coordinate singularity that can be removed to open up yet further regions. All these extensions are completely free of curvature singularities.

\subsection{Solution 3}
Finally, we intend to combine solutions 1 and 2 and embed them in a larger family. To this end, we solve the Ernst equation for the initial potential 
\begin{equation}
 t=0:\quad \E=c_1 (1-x^2)\left(1 - \frac{x}{d}\right) 
             + \ii x \left[c_3(x^2-3)-x)\right],
\end{equation}
which reduces to \eqref{eq:indat1} for $d\to\infty$, and to \eqref{eq:indat2} for $c_3\to0$. The resulting metric depends on the four parameters $c_1>0$, $c_3\in\R$, $d>1$ and $R_0>0$, and the potentials are given by
\begin{eqnarray}
 \ee^u &=&  8 c_1 d\frac{V}{(1 + y) (U^2 + V^2)},\quad
 \ee^M = \frac{d R_0W^2}{16c_1^3(d^2-1)^2r_d}\,(U^2 + V^2),\\
 Q     &=&  x + \frac{1-x^2}{4}
           \Bigg[2 c_3(2 + y) \nonumber\\
       && \qquad
        + \frac{1}{d+r_d-xy} \left(\frac{(1 + y)U}{c_1 d} 
                + \frac{c_3 (1-y)U^2}{2c_1^2V} - c_3 r_d(1-y^2)\right)\Bigg],\label{eq:formulaQ}
\end{eqnarray}
where
\begin{eqnarray}
 \hspace{-5mm}r_d &=& \sqrt{(d-xy)^2-(1-x^2)(1-y^2)},\quad
   U = 2 c_1 d\, [1-c_3(d-r_d+2x)](1-y),\\
 \hspace{-5mm}V &=& c_1^2 (d+r_d-xy)(1+y) 
      + c_3^2d^2\frac{(1-x^2)(1 - y)^3}{d+r_d-xy},\quad
    W = \frac{r_d+x-dy}{1-y}.
\end{eqnarray}
As for ``solution 1'' we have singularities for $c_3=\pm1$. The causal structure, on the other hand, is the same as for ``solution 2''. Here, however, the extensions are not necessarily regular and, depending on the parameter values, may or may not contain curvature singularities. Details will be presented in a future work\cite{Hennig}.

\section{A Family of Electrovacuum Solutions}

Since also the coupled system of the Einstein-Maxwell equations with Gowdy symmetry has an Ernst formulation (with two complex potentials $\E$ and $\Phi$, where $\E$ essentially encodes the metric and $\Phi$ the electromagnetic field), there is some hope to construct exact solutions in this case too.
These could provide interesting examples of cosmological models with a non-trivial additional field.

In order to construct solutions, instead of again solving an initial value problem, we can make use of our above solutions and apply a Harrison transformation\cite{Harrison1968}, which maps vacuum into electrovacuum solutions. To this end, we start from the Ernst potential $\E$ of ``solution~1'' and construct, for $\gamma\in\R$, an electrovacuum solution $\E'$, $\Phi'$ via
\begin{equation}
  \E'=\frac{\E}{1+\gamma^2\E},\quad
  \Phi'=\frac{\gamma\E}{1+\gamma^2\E}.
 \end{equation}
It is possible to compute the corresponding metric and the electromagnetic four-potential $(A_i)=(0,0,A_3,A_4)$ exactly, but the expressions are very lengthy. Instead we only give the first non-trivial correction to the case $\gamma=0$ (which is ``solution 1''):
\begin{eqnarray}
  \hspace{-5mm}
  \ee^u &=& \frac{16c_1 U}{(1 + y)(U^2 + V^2)}\times
           \nonumber\\
   \hspace{-5mm}&&\ \times\Bigg[1 +2\gamma^2\left((1-x^2)(1+y)\frac{U^2 + V^2}{16c_1U} 
                -(1-y)\frac{16c_1 U}{U^2+V^2}\tilde Q^2\right)\!\!\Bigg] 
             +\mathcal O(\gamma^4),\\
 \hspace{-5mm} Q &=& \left[1 - \gamma^2(1-x^2)(1+y)\frac{U^2+V^2}{4c_1 U}\right]\tilde Q
         +\mathcal O(\gamma^4),\\
  \hspace{-5mm}\ee^M &=&  \frac{R_0(U^2 + V^2)}{64c_1^3}
     \times \nonumber\\
   \hspace{-5mm}&& \ \times\Bigg[1+ 2 \gamma^2\left((1-x^2)(1+y)\frac{U^2+V^2}{16c_1U}
                + (1-y)\frac{16c_1U}{U^2+V^2}\tilde Q^2\right)\!\!\Bigg]
             +\mathcal O(\gamma^4),\\
  \hspace{-5mm}A_3 &=& \gamma\sqrt{R_0} (1-y)\frac{16c_1U}{U^2+V^2}\tilde Q 
         +\mathcal O(\gamma^3),\\
  \hspace{-5mm}A_4 &=& \gamma\sqrt{R_0}\left[(1-x^2)(1+y)\frac{U^2+V^2}{16c_1U}
             + (1-y)\frac{16c_1U}{U^2 + V^2}\tilde Q^2\right]
              +\mathcal O(\gamma^3),
\end{eqnarray}
 where
 \begin{eqnarray}
  U &=& 4c_1^2(1+y) + c_3^2(1-x^2)(1-y)^3,\quad
  V = 4c_1(1-y)[1-c_3 x(2+y)],\\
  \tilde Q &=& x + \frac{c_3}{8}(1-x^2)\left(7+4y+y^2 
    + \frac{(1-y)V^2}{4c_1^2U}\right).
 \end{eqnarray}
This solution has singularities for $c_3=\pm1$, but is otherwise regular in the Gowdy square, and the boundaries $t=0,\pi$ are Cauchy horizons. Again it is possible to construct extensions, and the causal structure turns out to be the same as for ``solution 1''. 

An important conclusion is that the Cauchy horizons in ``solution 1'' are stable under particular electromagnetic perturbations, namely the 1-parameter family of  perturbations described by the parameter $\gamma$ in ``solution 3''.

A detailed discussion of this electrovacuum solution will be presented elsewhere.\cite{Hennig}

\section*{Acknowledgments}
 I would like to thank Gerrard Liddell for commenting on the manuscript. 
This work was supported by the Marsden Fund Council from Government funding, administered by the Royal Society of New Zealand.


\end{document}